\documentclass[amsmath,amssymb,
               aps,prl,reprint,superscriptaddress,%
               showpacs,showkeys,lengthcheck%
              ]{revtex4-1}

\usepackage{bm}
\usepackage{type1cm}
\usepackage[bookmarks]{hyperref}
\usepackage{color}
\usepackage{graphicx}

\renewcommand{\Im}{\mathop{\mathrm{Im}}\nolimits}

\def\Xint#1{\mathchoice%
   {\XXint\displaystyle\textstyle{#1}}%
   {\XXint\textstyle\scriptstyle{#1}}%
   {\XXint\scriptstyle\scriptscriptstyle{#1}}%
   {\XXint\scriptscriptstyle\scriptscriptstyle{#1}}%
   \!\int}
\def\XXint#1#2#3{{\setbox0=\hbox{$#1{#2#3}{\int}$}
     \vcenter{\hbox{$#2#3$}}\kern-.5\wd0}}
\def\ddashint{\Xint=}

\def\clap#1{\hbox to 0pt{\hss#1\hss}}

\def\mathclap{\mathpalette\mathclapinternal}

\def\mathclapinternal#1#2{%
\clap{$\mathsurround=0pt#1{#2}$}}

\begin{document}
%\preprint{AAA/123-BI}

\title{Trace formula for counting nodal domains on the boundaries of chaotic 2D
billiards}

\author{Amit Aronovitch}
\email{amit.aronovitch@weizmann.ac.il}
\affiliation{Department of Physics of Complex Systems, %
The Weizmann Institute of Science, 76100 Rehovot, Israel}
\author{Uzy Smilansky}
\affiliation{Department of Physics of Complex Systems, %
The Weizmann Institute of Science, 76100 Rehovot, Israel}
\affiliation{School of Mathematics, Cardiff University, Wales UK}

%\date{\today}

\begin{abstract}
Given a Dirichlet eigenfunction of a 2D quantum billiard, the
boundary domain count is the number of intersections of the nodal
lines with the boundary. We study the integer sequence defined by
these numbers, sorted according to the energies of the
eigenfunctions. Based on a variant of Berry's random wave model, we
derive a semi-classical trace formula for the sequence of boundary
domain counts. The formula consists of a Weyl-like smooth part, and
an oscillating part which depends on classical periodic orbits and
their geometry. The predictions of this trace formula are supported
by numerical data computed for the Africa billiard.
\end{abstract}

%\keywords{}

\maketitle

Recently, the study of nodal patterns witnessed a remarkable
renaissance, and  attracted  the active interest of scientists from
very diverse fields---quantum chaos, acoustics, optics, spectral
theory, percolation and more~\cite{Wittenberg07}. The number of
nodal domains---connected components on which a real wave unction
has a constant sign, is an important feature, which has been used to
characterize eigenfunctions of wave equations and even to resolve
iso-spectral ambiguities~\cite{Gnutzmann05}. For quantum billiards,
the number $\nu_n$ of nodal domains of the $n$'th eigenfunction
(sorted by increasing eigenvalues $E_n\le E_{n+1}$) can never exceed
$n$~\cite{Courant}. The number of nodal domains can be computed
explicitly for separable systems. However, in the non-separable
cases, there exists no analytical tool which provides the number of
nodal domains, and even the numerical counting problem is difficult
due to the dependence on the detailed structure of the domains.
In~\cite{Blum} it was shown that the asymptotic distribution of
$\nu_n$ depends on the dynamics of the underlying classical system.
In the separable case, the parameters of the classical phase space
determine the nodal counts in the semi-classical limit. In the
chaotic case, the distribution matches the predictions of a
percolation model, which was proposed (but not rigorously justified)
in~\cite{Bogomolny02}.

The partition to nodal domains induces a partition of the boundary
to ``boundary domains''. They are defined as the nodal domains of
a``boundary function'', which for Dirichlet billiards, is the normal
derivative of the eigenfunction on the boundary. In 2D, the number
$\eta_n$ of boundary domains equals the number of nodal points
separating them---the boundary intersection (BI) points. This number
is more accessible, both numerically and theoretically, than
$\nu_n$. At the same time, it carries the fingerprints of the
underlying classical dynamics of the billiard~\cite{Blum}. Also, the
number of boundary domains provides information which is essential
for estimating the total number of nodal domains~\cite{Polterovich}.
It was recently shown~\cite{Toth07} that $\eta_n=O(\sqrt{n})$.
In~\cite{Blum}, the asymptotic distribution of $\eta_n/\sqrt{n}$ was
computed for a class of integrable billiards. For a chaotic billiard
of area $\mathcal{A}$ and boundary length $\mathcal{L}$, a random
wave model yields the estimate~\cite{Blum,Aronovitch} $\eta_n\sim
\mathcal{L}q/(2\pi)$, where $q=\sqrt{4\pi n/\mathcal{A}}$ is the
leading asymptotic estimate for the $n$'th wave-number
$k_n=\sqrt{E_n}$.

The purpose of the present paper is to go beyond the simple estimate
$\eta_n\sim \mathcal{L}q/(2\pi)$, and provide a trace formula which
approximates the mean value as well as the fluctuations in the
sequence $\eta_n$, in terms of the periodic orbits of the classical
billiard. Such formulae were proposed in the past for the counting
of nodal domains in separable billiards~\cite{Gnutzmann06}. Here is the
first time that a counting trace formula is written down for the
chaotic case.

% The derivation described in this letter uses a variant of that
%model to extend this result to a detailed trace formula, accounting
%for the oscillations around the asymptotic $O(\sqrt{n})$ mean, in
%terms of the periodic orbits of the underlying classical billiard.

The counting of boundary intersections is performed by computing the
density $d_\eta(n)=\sum_{m\in \mathbb{N}^*} \delta(n-m)\eta_m$. For
a chaotic Dirichlet billiard with a smooth boundary, we derive in
the sequel the following asymptotic expression:
\begin{eqnarray}
\hspace*{-5mm} d_\eta(n)&\approx& \frac{\mathcal{L}}{2\pi}q
       +\frac{\mathcal{L}^2 - 6\pi\mathcal{A}}{4\pi\mathcal{A}} \nonumber\\*
\hspace*{-5mm} &+&\frac{1}{\pi}
           \sum_{p,r}
              \frac{\Phi_p }
                   {\sqrt{|\mathrm{tr}[{M_p}^{r}-\mathrm{I}]|}}
              \cos(r(\tilde{q}L_{p}-\nu_{p}\tfrac{\pi}{2})),
\label{eq:bi-tf-n}
\end{eqnarray}
where $p$ enumerates classical periodic orbits, $r\in \mathbb{N}^*$ counts
repetitions of the orbit, $L_p$ is the length of the orbit, $M_p$ the monodromy
matrix, $\nu_p$ the Maslov index, and
$\tilde{q}= q+\mathcal{L}/(2\mathcal{A})$. $\Phi_p$ is a trigonometrical factor
depending on the $n_p$ bounce angles of the orbit $p$:
\begin{equation}
\Phi_p = \sum_{i=1}^{n_{p}}
        (4\cos^{2}\psi_{i}^{(p)}-1)2 \sin\psi_{i}^{(p)}.
\label{eq:bi-trigfact}
\end{equation}

Before sketching the derivation of the  trace formula, we shall
demonstrate its application. We computed the lowest 20,000
eigenfunctions ($0 < k_n < 260 $) of the Africa billiard~\cite{Berry86},
the corresponding BI count sequence $\eta_n$, and
the density $d_\eta^\rho(q)=\sum_n\rho(q-\sqrt{4\pi
n/\mathcal{A}})\eta_n$ (where here and in what follows $\rho(x)$ is
a narrow Gaussian approximating the Dirac $\delta$. $\rho$ is
defined as a density, so
$\rho(q-q_0)=\rho(n-n_0)\cdot\mathcal{A}q_0/(2\pi)$). Subtracting
the predicted smooth part $d_\eta^{\mathrm{sm}}(q)$ and scaling, we
computed $f(q)=(d_\eta^\rho -d_\eta^{\mathrm{sm}})/q\cdot W(q)$, where
$W$ is a Gaussian ``window function'' of width $\sigma = 50$ and
center $q_0=130$, which was used for softening the sharp cutoff due
to the finiteness of the computed spectrum. The length spectrum,
which is the Fourier transform $\hat{f}(x)$ of $f(q)$ was compared
with $\hat{f}_{\mathrm{scl}}(x)$, the theoretical prediction based on the
oscillating part of~(\ref{eq:bi-tf-n}). The latter was computed
using 70 classical periodic orbits (with up-to 7 bounce points) and
15 complex periodic orbits whose lengths had a very small imaginary part
(however, the complex orbits did not have a significant effect).
\begin{figure*}
  \includegraphics[clip,scale=0.75]{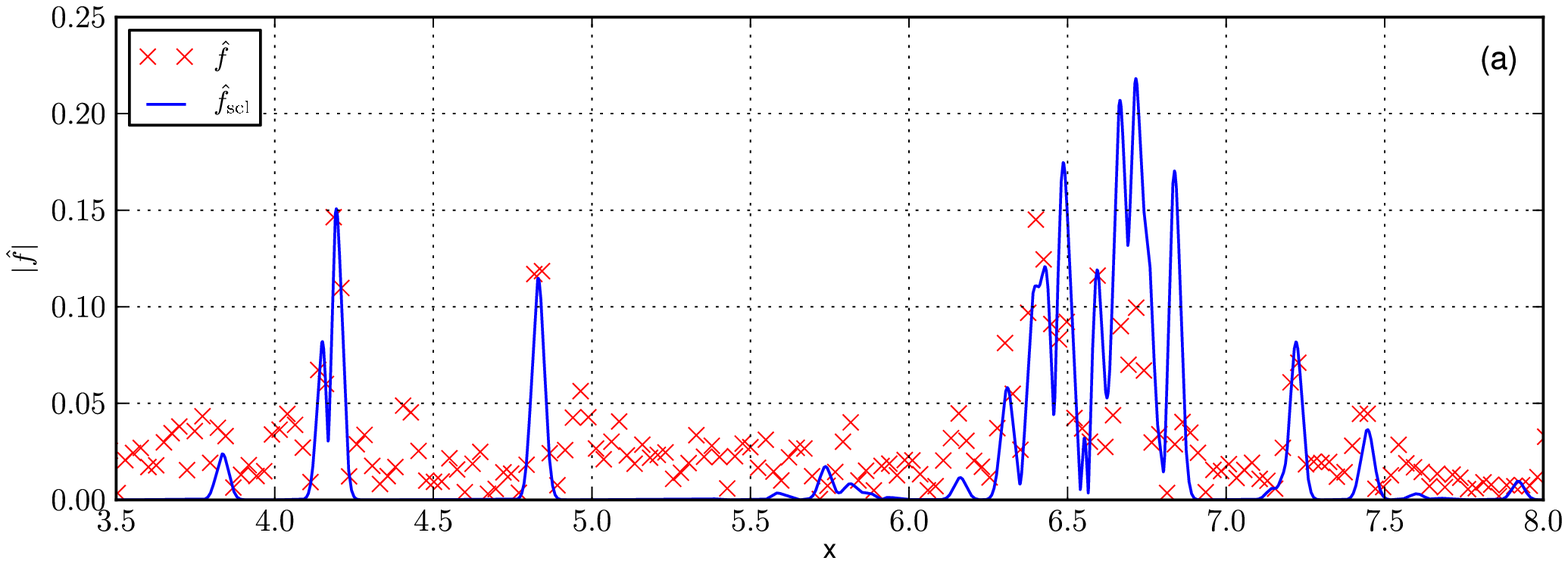}
  \includegraphics[clip,scale=0.75]{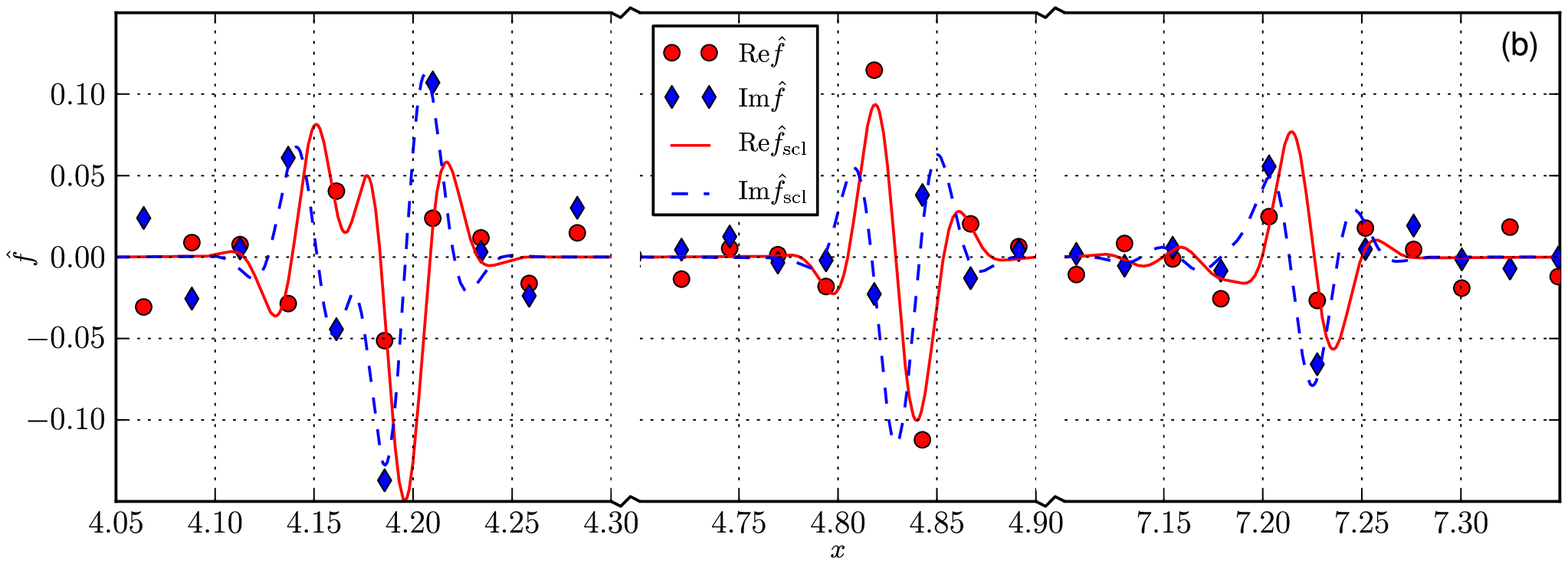}
  \caption{\label{cap:dbiq-tf}%
   The semi-classical and numerical length spectra (Fourier transform of
   $d^{\rho}_{\eta}(q)$).
   (a): Absolute value, (b): Magnified view of 3 prominent peaks.}
\end{figure*}

FIG.~\ref{cap:dbiq-tf}(a) displays several peaks centered at lengths
of periodic orbits which match quite well with the theoretical
predictions. A more detailed comparison is presented in
FIG.~\ref{cap:dbiq-tf}(b). Due to~(\ref{eq:bi-trigfact}), orbits
whose angles are close to 60$^\circ$ are inhibited. Indeed, the
triangular periodic orbits of the billiard, whose lengths are in the
range 5.07--6.05, cannot be seen above the background level. The
structure around $x=6.5$ is due to several periodic orbits that pass
very close to the boundary at the region of its highest concavity.
The poor agreement between the semi-classical theory and the
numerics in this region is due to penumbra
corrections~\cite{Primack96} which were not included. The random
background of amplitude $\sim 0.04$ observed in the plots  does not
seem to diminish when $q_0$ increases (within the range of our
study). This phenomenon will be discussed below.

The derivation of the trace formula~(\ref{eq:bi-tf-n}) starts by
expressing $\eta_n$ as $\eta_n=\oint b_n(s)ds$, where $0\le s<\mathcal{L}$
is the boundary arc length and the BI density $b_n(s)$ is given by
\begin{equation}
b_{n}(s) = \sum_{i=1}^{\eta_n} \delta(s_i^{(n)}-s)
        = \delta\!\left(u_{n}(s)\right)|\dot{u}_{n}(s)| .
\label{eq:bidenst}
\end{equation}
Here, $u_n(s)=\bm{n}(s)\cdot
\boldsymbol{\nabla}\psi_n(\bm{r}(s))/k_n$ is the scaled normal
derivative of the $n$'th eigenfunction $\psi_n$, taken at the
boundary point $\bm{r}(s)$,  $\dot{u}_{n}(s)=\frac{d u_n(s)}{ds}$,
and $s_i^{(n)}$ are the BI points (zeros
of $u_{n}$).

The trace formula will be derived for a smoothed version
$d_\eta^\rho(n)=\int\! d_\eta (m)\rho(n-m)\, dm =
 \sum_{m\in\mathbb{N}^*} \rho(n-m)\eta_m$, (The smoothing kernel $\rho$ was
defined above). In the sequel we shall consistently use the notation
$d_X^\rho(y)$ to denote the $\rho$ smoothed density of the quantity
$X$ in the variable $y$.

Consider the spectral density of $b(s)$ at wavenumber $k$,
$d_b(s;k)=\sum_n\delta(k_n-k)b_n(s)$. It is approximated by
$d_b^\rho(s;k)=d^\rho(k)\langle b_n(s) \rangle_{n}$, where
$d^\rho(k)=\sum_n \rho(k_n-k)$ is the smoothed spectral density and
$\langle b(s) \rangle_{n}=\sum_n  b_n(s)\cdot \rho(k_n-k)/d^\rho(k)$
is the spectral average of $b_n(s)$ around $k$. If we choose the
width of $\rho$ to be of order $ k^{-\frac{1}{2}} $, then as
$k\rightarrow\infty$, the discrete ensemble of boundary functions in
the corresponding spectral window around $k$ approaches a continuous
distribution. To proceed, we introduce the conjecture that for
chaotic billiards this limiting distribution is Gaussian. This
conjecture, may be seen as a variant of Berry's random wave
model~\cite{Berry77}, adjusted to the boundary as in~\cite{Berry02}
and~\cite{Wheeler}. While the validity of the conjecture is expected
to improve in the semiclassical limit, numerical tests within the
range of $k$ values used here reveal a residual Kurtosis which might
explain the background observed in the length spectrum. Adopting
the random waves conjecture enables us to express the mean of the
density $b(s)$ in terms of the variances of the field $u(s)$ and its
derivative $\dot{u}(s)$~\cite{Rice45}:
\begin{equation}
\langle b(s) \rangle =
      \frac{1}{\pi} \sqrt{\frac{\langle \dot{u}^2(s)\rangle}
                               {\langle u^2(s)\rangle } }
     = \frac{1}{\pi} \sqrt{\frac{d^\rho_{\dot{u}^2}(k)}
                                {d^\rho_{u^2}(k)} }\ .
\label{eq:rice}
\end{equation}
To compute the required densities, we write
\begin{equation}
d_{u^2}(s)=\sum_n \delta(k_n-k){u_n}^2(s)=\frac{2k}{\pi}\Im g(s,s;k)
\label{eq:denst-uu}
\end{equation}
where $g(s,s';k)=\sum_n u_n(s)u_n(s')/({k_n}^2-k^2)$ is the boundary
Green function. As shown in~\cite{Backer}, $g$ can be expanded as
$g=\sum_{n=0}^\infty \hat{h}^n g_0$, where $\hat{h}$ is the integral
operator with kernel function $h$. The functions $g_0$ and $h$ are given by
\begin{eqnarray*}
h(s,s';k)&=&2\bm{n}(s)\cdot
     \boldsymbol{\nabla}_{\bm{r}(s)}G_{0}\!\left(\bm{r}(s), \bm{r}(s') ; k
                                            \right) \\
g_{0}(s,s';k)&=&\frac{2}{k^{2}}\sum_{i,j}n_{i}{n'}_{j}
              \frac{\partial^{2}G_{0}\!\left(\bm{r}(s), \bm{r}(s'); k\right)}
                   {\partial r_{i}\partial r'_{j}},
\end{eqnarray*}
and $G_0 = \frac{i}{4}H_0^+(k|\bm{r}-\bm{r}'|)$ is the free Green
function in 2D. To handle the singularities involved in this
expansion and make it more amenable to semi-classical treatment, we
choose a large cutoff $1\ll x_C\ll k\mathcal L$, and split $g_0$
into a ``far'' (off diagonal) part and a ``near'' (close to the
diagonal) part:
\begin{eqnarray*}
g_{0}^{(N)} &=& g_{0}(s,s') H(x_C/k - |s-s'|), \\
g_{0}^{(F)} &=& g_{0}(s,s') H(|s-s'| - x_C/k ) \\
\end{eqnarray*}
(where $H$ is the Heaviside step function).
It can be shown that for large $k$
\begin{equation*}
\hat{h}g_0^{(N)} = \ddashint_0^{\mathcal{L}}h(s,s_1)g_0^{(N)}(s_1,s') ds_1 \\
       \sim  g_0^{(F)} + g_1^{(N)},
\end{equation*}
where $\ddashint$ denotes Hadamard finite part integration~\cite{Hadamard},
the near diagonal part $g_1^{(N)}$ is bounded and $\hat{h}g_1^{(N)}$ is
negligible.
Hence, the expansion of $g$ can be rewritten as:
\begin{equation}
g \sim g_0^{(N)}+ g_1^{(N)} + 2\sum_{n=0}^{\infty}{\hat{h}}^n g_0^{(F)}.
\label{eq:g_expand}
\end{equation}
Substituting this result in~(\ref{eq:denst-uu}), we get a similar expansion
for $d_{uu'}(s,s')$.
The first two terms, which were explicitly computed
in~\cite{Backer} for $s'\rightarrow s$, yield the ``smooth part''
$(k-\kappa(s))/(2\pi)$, where $\kappa$ is the curvature
(note that this can also be derived by applying the methods
of~\cite{Aronovitch} on the curved boundary corrections to Berry's
random wave model, which are described in~\cite{Wheeler}).
The third term is interpreted as a summation over possible paths
from $s'$ to $s$, where the $n$th summand includes integration over
$n$ intermediate bounce points. Approximating the integrals by the
stationary phase method (closely following~\cite{Smilansky96}), we
get a sum over \emph{classical\/} paths (allowing only specular
bounces). The oscillating part for $d_{uu'}$ is given by
\begin{equation}
    {\left(\frac{2}{\pi}\right)}^{\mathclap{\frac{3}{2}}}
    \sum_t
         \sqrt{\left|
                  \frac{\sin\psi \sin{\psi}'}
                       {{\left( \partial s /
                                \partial p' \right)}_{s'} }
               \right| }
          \cos(k L_t+\tfrac{3}{4}\pi-\tfrac{\pi}{2}\nu_{t}),
\label{eq:cosc-tf}
\end{equation}
where $t$ enumerates classical orbits from $s'$ to $s$, $L_t$ is the
length of the orbit, $\nu_t$ is the Maslov index (number of
conjugate points plus twice the number of bounce points), $\psi$ and
$\psi'$ are the angles between the orbit and the boundary at $s$ and
$s'$ respectively, and $p'=k\cos(\psi')$ is the classical momentum
at $s'$ (hence, the oscillating part of $d_{uu'}$ is $O(\sqrt{k})$).
Taking derivatives of this expansion, we get an expression for
$d_{\dot{u}\dot{u}'}=\partial_s\partial_{s'}d_{uu'}$ , which has a
similar form. Due to the rapid decay of the Fourier transformed convolution
kernel $\hat{\rho}$, the expansions (taken at $s=s'$) of $d^\rho_{uu'}$ and
$d^\rho_{\dot{u}\dot{u}'}$ converge, so the quotient required for substitution
in~(\ref{eq:rice}) can be computed, and used to derive an expression for
$\langle b(s)\rangle$. To compute $d_\eta^\rho(k)$, we multiply by $d^\rho(k)$
(for which we can use the Gutzwiller trace formula, again smoothed by
$\rho$ to ensure convergence), and integrate over $s$. The
stationary phase condition added by this extra integration ensures
that $\psi=\psi'$, and the result includes summation over
\emph{periodic\/} classical orbits. Finally, we want to discard the
spectral information and compute $\eta(n)$ rather than $\eta(k)$.
Therefore, following the method described in~\cite{Gnutzmann06}, we
substitute in the resulting expression for $d_\eta^\rho(k)$, an
expansion for $k(n)$, achieved by formally inverting the Gutzwiller trace
formula. This leads us to the trace formula for $d_\eta^\rho(n)$,
presented in~(\ref{eq:bi-tf-n}).

\begin{figure*}
  \includegraphics[clip,scale=0.75]{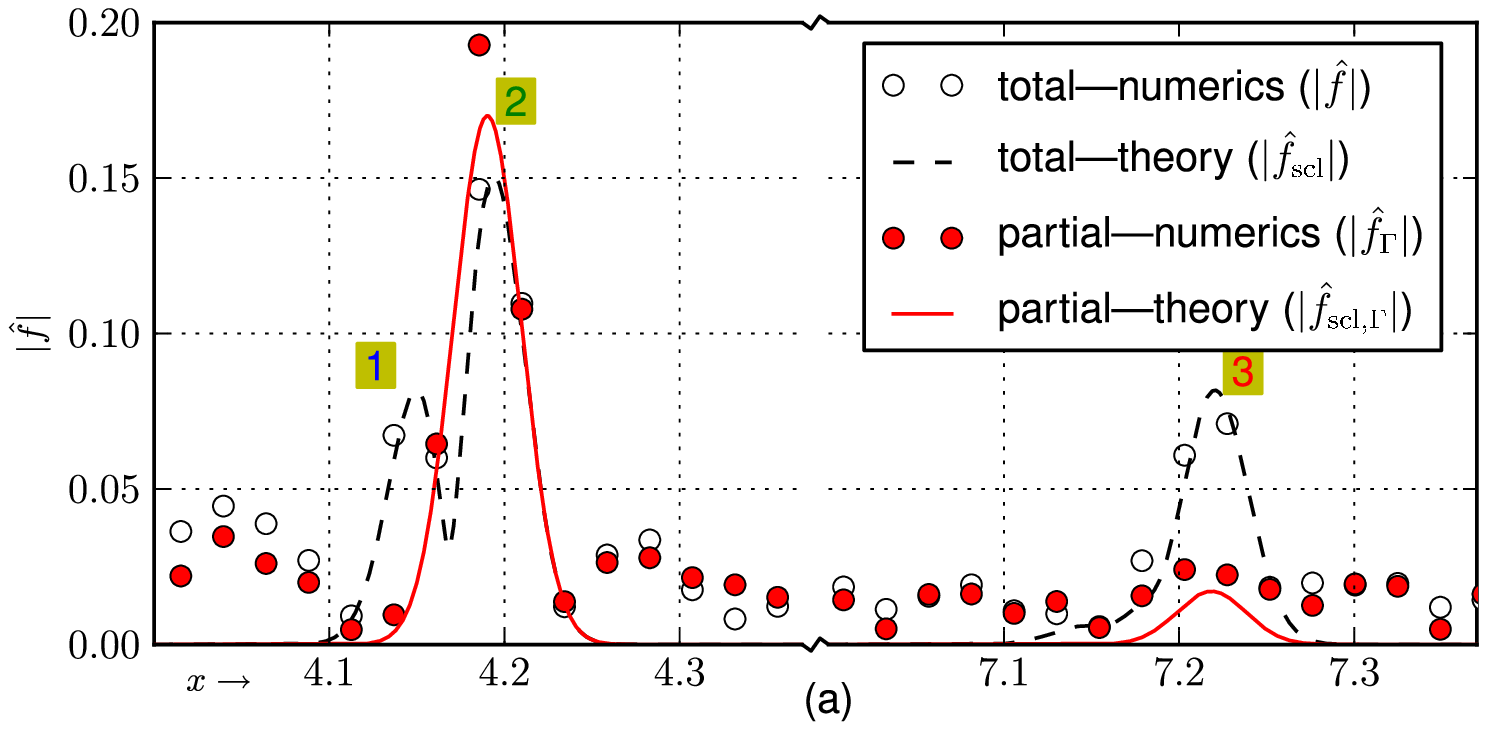}
  \includegraphics[clip,scale=0.75]{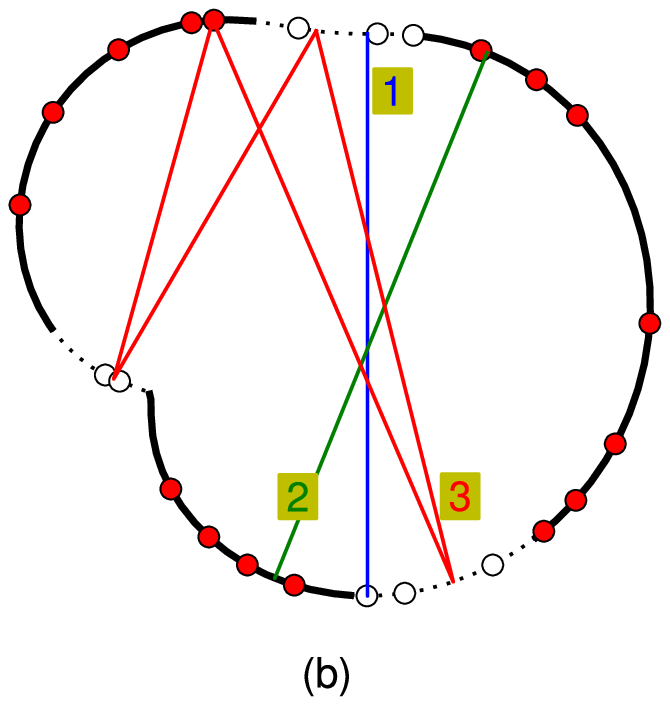}
  \caption{\label{cap:dbiq_exc}%
  Restricting the BIC to $\Gamma\subset\partial\Omega$ reduces %
  the amplitudes for orbits hitting the excluded region. Compare $\hat{f}$ %
  to $\hat{f}_\Gamma$ for the 3 marked orbits.}%
\end{figure*}

A stringent test of the theory above, is based on the following
argument. A ``partial'' trace formula which counts BI located on a
prescribed part $\Gamma\subset \partial\Omega$ of the boundary can
be similarly derived by integrating the BI density over $\Gamma$
alone: $d_{\eta\,\Gamma}(k)=\int_\Gamma d_b(s;k) ds$. Since the
formula for $d_b$, much like~(\ref{eq:cosc-tf}), involves summation
over orbits starting and ending at $s$, we conclude that only
periodic orbits that have a bounce point in $\Gamma$ will contribute
to the sum in the resulting trace formula for $d_{\eta\,\Gamma}$. By
choosing a $\Gamma$ which is bounded away from the bounce points of
a specific orbit, we can effectively turn off the effect of that
orbit. Similarly, we expect orbits that have some, but not all of
their bounce points in the excluded regions $\partial\Omega\setminus
\Gamma$, to have reduced amplitude in the length spectrum. This
result is demonstrated in FIG.~\ref{cap:dbiq_exc}.
In FIG.~\ref{cap:dbiq_exc}(b), $\Gamma$ is plotted with a wide line,
while the excluded part $\partial\Omega\setminus \Gamma$ is dotted.
For demonstration purposes, the BI of $\psi_{150}$ are shown on the
boundary (total $\eta_{150}=24$), and the points to be excluded from
$\eta_\Gamma$ are marked with empty circles ($\eta_{\Gamma\,
150}=16$). Three orbits are shown, and the corresponding peaks in
the length spectrum are also marked in FIG.~\ref{cap:dbiq_exc}(a).
Orbit 1 has both its bounce points in the excluded regions, so it
completely disappears from the length spectrum corresponding to the
partial count $\eta_\Gamma$. Orbit 2, which has both of its bounce
points in $\Gamma$ is not effected by the exclusion, and orbit 3,
which has only 1 out of 4 bounce points in $\Gamma$, is
significantly inhibited, and drops below the noise level for the
numerical case. This test and the general agreement between the
semi-classical and the numerical length spectra give credence to the
validity of the proposed trace formula.

\begin{acknowledgments}
We would like to thank Klaus Hornberger for his assistance in
deriving the required hypersingular operators, and Roman Schubert
for valuable discussions.
The research was supported by the Minerva and Einstein (Minerva) Centers
at the Weizmann Institute, and by grants from NSF (grant 2006065),
ISF (grant 166/09), and EPSRC (grant GR/T06872/01.)
\end{acknowledgments}

\bibliography{bi_chaotic}

\end{document}